\documentclass{article}

\usepackage{eusipco2009}                       
\usepackage{graphicx}

\title{Eigenresiduals for improved Parametric Speech Synthesis}

\name{Thomas Drugman $^1$, Geoffrey Wilfart $^2$, Thierry Dutoit $^1$}

\address{$^1$ TCTS Lab, Facult\'e Polytechnique de Mons - 31, Boulevard Dolez, 7000, Mons, Belgium \\
$^2$ Research and Development, Acapela Group - 33, Boulevard Dolez, 7000, Mons, Belgium \\
phone: + (32) 65 37 47 49, fax: + (32) 65 37 47 29\\ email:
thomas.drugman@fpms.ac.be \\
}

\begin{document}

\maketitle

\begin{abstract}
Statistical parametric speech synthesizers have recently shown their ability to produce natural-sounding and flexible voices. Unfortunately the delivered quality suffers from a typical \emph{buzziness} due to the fact that speech is vocoded. This paper proposes a new excitation model in order to reduce this undesirable effect. This model is based on the decomposition of pitch-synchronous residual frames on an orthonormal basis obtained by Principal Component Analysis. This basis contains a limited number of \emph{eigenresiduals} and is computed on a relatively small speech database. A stream of PCA-based coefficients is added to our HMM-based synthesizer and allows to generate the voiced excitation during the synthesis. An improvement compared to the traditional excitation is reported while the synthesis engine footprint remains under about 1Mb. 

\end{abstract}


\section{Introduction}\label{intro}

For the last decade, Unit Selection-based methods \cite{Hunt} have clearly emerged in speech synthesis. These techniques rely on a huge corpus (typically several hundreds of Mb) covering as much as possible the diversity one can find in the speech signal. During synthesis, speech is obtained by concatenating natural units picked up from the corpus. As the database contains several examples for each speech unit, the problem consists in finding the best path through a lattice of potential candidates by minimizing a selection and concatenation cost. This approach generally generates speech with high naturalness and intelligibility. However quality may degrade severely when an under-represented unit is required or when a bad jointure (between two selected units) causes a discontinuity. 
 
More recently, a new synthesis method has been proposed: the Statistical Parametric Speech Synthesis \cite{Black}. This approach relies on a statistical modeling of speech parameters. After a training step, it is expected that this modeling has the ability to generate realistic sequences of such parameters. The most famous technique derived from this framework is certainly the HMM-based speech synthesis \cite{Tokuda}, which obtained in recent subjective tests a performance comparable to Unit Selection-based systems \cite{Blizzard}. An important advantage of such a technique is its flexibility for controling speech variations (such as emotions or expressivity) and for easily creating new voices (via statistical voice conversion). Its two main drawbacks, due to its inherent nature, are:

\begin{itemize}
\item the lack of naturalness of the generated trajectories. The statistical processing tends to remove details in the feature evolution; generated trajectories are oversmoothed, which makes the synthetic speech sound muffled. Some approaches considering global variance \cite{GV} or trajectory HMMs \cite{Trajectory} have been proposed in order to reduce this detrimental effect. 
\item the "buzziness" of produced speech, which suffers from a typical vocoder quality. 
\end{itemize}

\begin{figure}[!ht]
  \centering
  \includegraphics[width=0.4\textwidth]{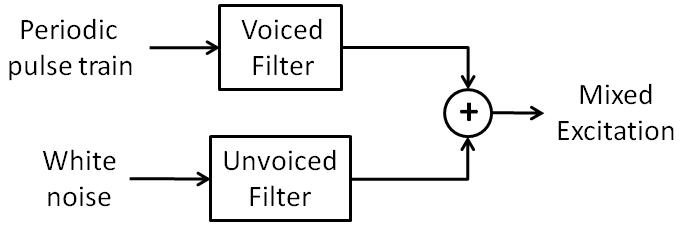}
  \caption{Mixed Excitation workflow.}
  \label{fig:ME}
\end{figure}

While the parameters characterizing spectrum and prosody are rather well-established, improvement can be expected by adopting a more suited excitation modeling. Indeed the traditional excitation considers either a white noise or a pulse train during unvoiced or voiced segments respectively. Inspired from the physiological process of phonation where the glottal signal is composed of a combination of periodic and aperiodic components, the use of a Mixed Excitation (ME) has been proposed. The ME is generally achieved as in Figure \ref{fig:ME}. In \cite{Yoshimura}, the filter coefficients were derived from bandpass voicing strenghts. In \cite{Maia}, state-dependent high-degree filters were directly trained using a closed loop procedure. The integration of a Liljencrants-Fant waveform as a modeling of the glottal source, possibly producing different voice qualities by varying the LF parameters, was proposed in \cite{Cabral2}. In \cite{Drugman}, we suggested the use of a codebook of typical pitch-synchronous residual frames. All these approaches reported a certain improvement with regard to the traditional excitation. 

\begin{figure}[!ht]
  \centering
  \includegraphics[width=0.45\textwidth]{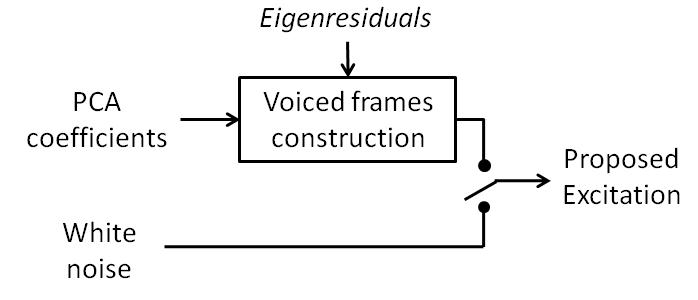}
  \caption{The proposed excitation workflow.}
  \label{fig:ProposedExcitation}
\end{figure}

This paper proposes a new excitation model in order to reduce the buzziness of parametric speech synthesizers. This model is based on the decomposition of pitch-synchronous residual frames on an orthonormal basis obtained by Principal Component Analysis (PCA). This basis contains a limited number of \emph{eigenresiduals} and is computed on a relatively small speech database ($\approx$ 20 min), from which a dataset of voiced frames is extracted. These frames have the particularity of being centered on a Glottal Closure Instant (GCI), two-period long and Hanning-windowed. Furthermore they are resampled on a fixed number of points and normalized in energy. This is required to ensure inter-frame coherence before applying PCA. Once the PCA transform is calculated, the whole corpus is analyzed and PCA-based parameters are extracted. This allows us to enhance our HMM-based speech synthesizer with a new stream of excitation parameters, besides the traditional pitch feature. During synthesis, voiced frames are derived from the generated PCA coefficients and overlap-added so as to obtain the excitation signal (Figure \ref{fig:ProposedExcitation}).

The paper is structured as follows. Section \ref{sec:Eigenres} describes the way the eigenresiduals are obtained from a speech database. For this, a dataset of normalized residual frames is extracted (\ref{ssec:dataset}) and PCA is calculated on it, allowing dimensionality reduction (\ref{ssec:PCA}). Section \ref{sec:HTS} details how this approach is integrated into an HMM-based speech synthesizer, relying on the framework available in \cite{HTS}, and presents the results of perceptual listening tests. Experiments throughout the paper were performed on three speakers: AWB (Scottish male) and SLT (US female) from the publicly available CMU ARCTIC speech database \cite{CMU}, and Bruno (French male) kindly provided by Acapela Group. Finally Section \ref{sec:conclu} concludes and proposes some guidelines for future works.

\section{Eigenresiduals through Speech Analysis}\label{sec:Eigenres}

The goal of this Section is to describe how eigenresiduals are obtained from a speech database. A dataset of normalized pitch-synchronous residual frames is first extracted (Section \ref{ssec:dataset}). As this set contains comparable data, Principal Component Analysis (PCA) can be calculated from it (Section \ref{ssec:PCA}) and a limited number of eigenresiduals can be retained.

\subsection{Obtaining normalized pitch-synchronous residual frames}\label{ssec:dataset}

\begin{figure}[!ht]
  \centering
  \includegraphics[width=0.45\textwidth]{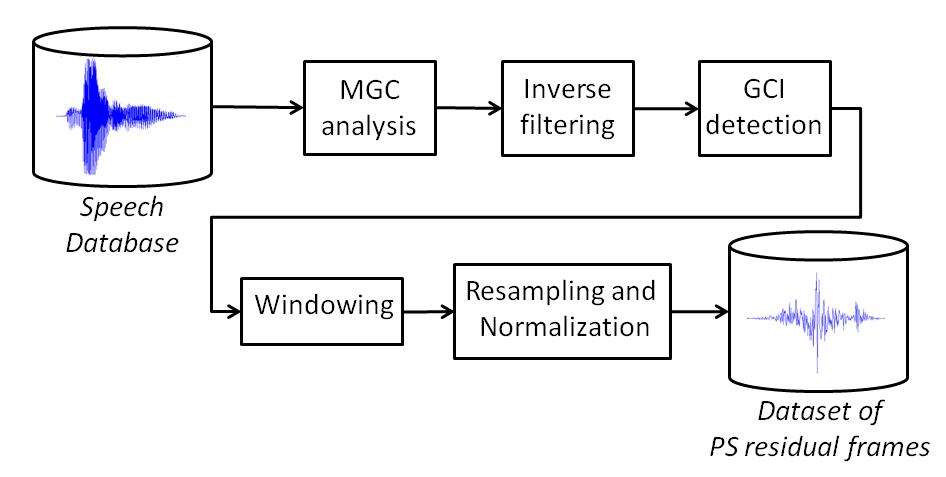}
  \caption{Obtaining a dataset of normalized pitch-synchronous residual frames from a speech database.}
  \label{fig:Dataset}
\end{figure}

\begin{figure}[!ht]
  \centering
  \includegraphics[width=0.45\textwidth]{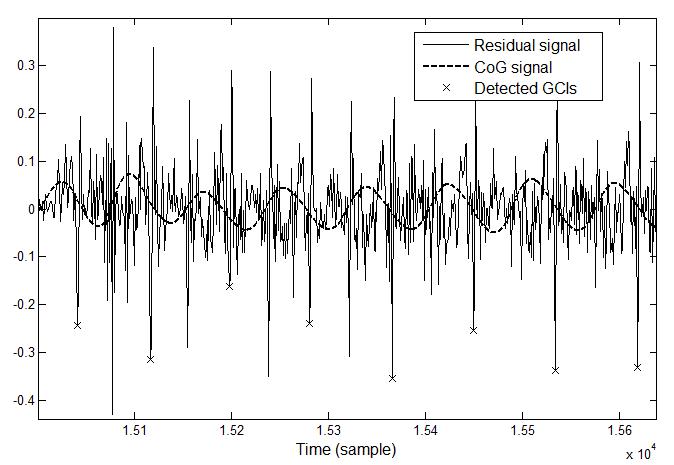}
  \caption{Determining the GCI location using the Center of Gravity technique.}
  \label{fig:GCI}
\end{figure}

Mel-Generalized Cepstral coefficients (MGC) have been designed so as to accurately and robustly capture the spectral envelope of speech
signals \cite{MGC}. The workflow presented in Figure \ref{fig:Dataset} thus performs MGC analysis with $\delta=0.42$ ($Fs=16kHz$) and $\gamma=-1/3$, as these values gave the best perceptual results in \cite{Blizzard}. Residual signals are then obtained by inverse filtering. As previously mentioned, an important characteristic of our residual frames is that they are centered on Glottal Closure Instants (GCIs). In order to locate GCIs, we use a method based on the Center of Gravity (CoG) in energy of the speech signal, as suggested in \cite{Kahawara}. While this gives straightforward results in most of cases, some residual segments are more contentious. Figure \ref{fig:GCI} exhibits how a peak-picking technique coupled with the detection of zero-crossings (from positive to negative) of the CoG signal removes possible ambiguities on the GCI positions.  

\begin{figure}[!ht]
  \centering
  \includegraphics[width=0.45\textwidth]{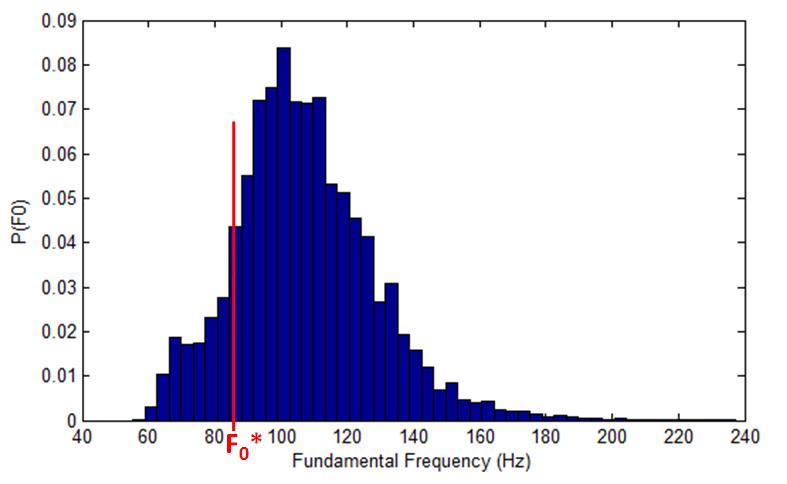}
  \caption{Analyzing a pitch histogram in order to determine the normalized pitch value (here for the male speaker Bruno).}
  \label{fig:PitchHisto}
\end{figure}

Residuals are then windowed by a two-period Hanning window. To ensure a point of comparison between residual frames before applying PCA, GCI-alignment is not sufficient. Indeed frames probably come from different prosodic contexts, so that some normalization in both pitch and energy is required. Pitch normalization is achieved by resampling, which retains the residual most important features. As a matter of fact, assuming that the residual obtained by inverse filtering approximates the glottal flow first derivative, resampling this signal preserves the \emph{open quotient}, \emph{asymmetry coefficient} (and consequently the $F_g$/$F_0$ ratio, where $F_g$ stands for the \emph{glottal formant} frequency) as well as the return phase characteristics (see \cite{Cabral}). Care has to be taken here when choosing the normalized pitch value, as this step will condition the synthesis quality. Indeed, at synthesis time, residual frames will be obtained by resampling a combination of eigenresiduals. If these have not a sufficiently low pitch, the ensuing upsampling will compress the spectrum and cause the appearance of "energy holes" at high frequencies. In order to avoid it, the speaker's pitch histogram $P(F_0)$ is analyzed and the normalized pitch value $F_0^*$ we chose typically satisfies:

\begin{equation}\label{eq:pitch}
\int_{F_0^*}^{\infty} P(F_0) \emph{d} F_0 \approx 0.8
\end{equation}

such that only $20\%$ frames will be slightly upsampled at synthesis time (see Figure \ref{fig:PitchHisto}). 

At this point, we have thus at our disposal a dataset of GCI-centered, Hanning-windowed, pitch and energy-normalized residual frames which is suited for applying PCA.  

\subsection{Eigenresiduals computation}\label{ssec:PCA}
Principal Component Analysis (PCA) is an orthogonal linear transformation which applies a rotation of the axis system so as to obtain the best representation of the input data, in the Least Squared (LS) sense \cite{PCA}. It can be shown that the LS criterion is equivalent to maximizing the data dispersion along the new axes. PCA can then be achieved by calculating the eigenvalues and eigenvectors of the data covariance matrix.

Let us assume that our dataset consists of $N$ residual frames of $m$ samples. PCA computation will lead to $m$ eigenvalues $\lambda_i$ with their corresponding eigenvectors $\mu_i$ (here called \emph{eigenresiduals}). $\lambda_i$ represents the data dispersion along axis $\mu_i$ and is consequently a measure of the "information" this eigenresidual conveys on the dataset. This is important in order to apply dimensionality reduction. Let us define $I(k)$, the "information rate" when using $k$ eigenresiduals, as the ratio of the dispersion along these $k$ axes over the total dispersion:

\begin{equation}\label{eq:info}
I(k)=\frac{\sum_{i=1}^k \lambda_i}{\sum_{i=1}^m \lambda_i}
\end{equation}

\begin{figure}[!ht]
  \centering
  \includegraphics[width=0.45\textwidth]{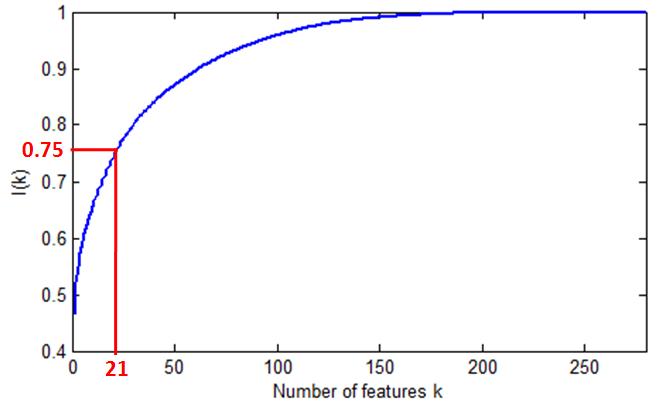}
  \caption{"Information" rate when using $k$ eigenresiduals for speaker AWB.}
  \label{fig:Pval}
\end{figure}

Figure \ref{fig:Pval} displays this variable for the male speaker AWB ($m=280$ in this case). Through subjective tests on an Analysis-Synthesis application, we observed that choosing $k$ such that $I(k)$ is greater than about 0.75 has almost inaudible effects when compared to the original file. Back to the example of Figure \ref{fig:Pval}, this implies that about 20 eigenresiduals can be efficiently used for this speaker.

\begin{figure}[!ht]
  \centering
  \includegraphics[width=0.45\textwidth]{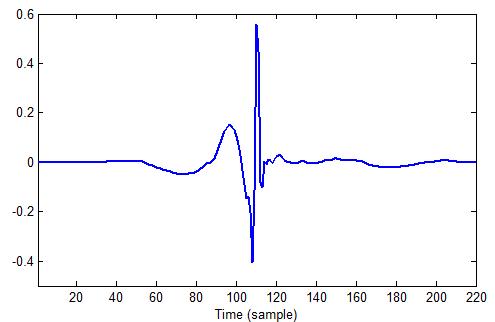}
  \caption{The first eigenresidual for the female speaker SLT.}
  \label{fig:EigRes}
\end{figure}

To give an idea of what an eigenresidual looks like, Figure \ref{fig:EigRes} exhibits the first eigenresidual, interpreted as the principal pattern arising from the data, for the female speaker SLT ($m=220$). A strong similarity with glottal flow models can be noticed, mainly during the glottal open phase.

\section{Eigenresiduals and Parametric Speech Synthesis}\label{sec:HTS}

\subsection{Our HMM-based speech synthesizer}

As previously mentioned, the HMM framework is very suitable in statistical parametric speech synthesis, as it has the ability to produce compact models. From a training perspective, it also uses well-known algorithms with guaranteed convergence. HMMs are built on the assumption that the speech signal is made of a concatenation of short quasi-stationary segments. Each segment corresponds to a state, and is viewed as a realisation of some statistical distribution, generally modeled by a Gaussian mixture. The training step may then look very close to building a speech recognizer, except that we use a context oriented clustering via binary decision trees.

%

Our feature vector consists of the 24-th order MGC parameters, log-$F_0$, and the PCA coefficients whose order has been determined as explained in Section \ref{ssec:PCA}, concatenated together with their first and second derivatives. Log-$F_0$ and PCA coefficients are however only meaningful in voiced segments of the signal. This leads to a modeling problem. On the one hand, some ``value'' needs to be provided for these parameters, even in unvoiced regions. On the other hand, providing a value in unvoiced regions will necessarily affect the Gaussian models.

In \cite{MSD}, authors propose an elegant solution and introduce the concept of Multi-Space Distribution (MSD). A MSD considers that the sample space is composed of several ``spaces'' that can have different dimensions (possibly 0). Each space has its own weight and, if the space dimension is non-zero, its probability distribution function (pdf). An event of dimension $n$ then consists of a set of $n$-dimensional spaces it belongs to, and its probability in each of these spaces. As shown in \cite{MSD}, MSD is a general framework including standard discrete and continuous density pdfs. Authors also show that it is suitable in the context of HMM modeling by deriving reestimation formulae.

For our purpose, we chose to model each of log-$F_0$ and PCA coefficients by 2-space distributions. One of the spaces corresponds to the voiced regions and has a diagonal-covariance single-Gaussian distribution. The other space for unvoiced regions, has zero-dimensionality, and is characterized only by its weight. The same model is adopted for first and second derivatives of the parameters. For the sake of simplicity, we chose to model static features, first and second derivatives of our MSD parameters as independent streams. Finally, we use a single-space diagonal-covaraiance single-Gaussian distribution for MGCs and their first and second derivatives. Our model therefore uses 7 streams, defined as follows:

\begin{itemize}
\item MGCs + $\Delta$MGCs + $\Delta$$\Delta$MGCs,
\item log-$F_0$,
\item $\Delta$log-$F_0$,
\item $\Delta$$\Delta$log-$F_0$,
\item PCAs,
\item $\Delta$PCAs,
\item and $\Delta$$\Delta$PCAs.
\end{itemize}

We use 5-state left-to-right context-dependent phoneme models, using pdfs described above. A state duration model is also determined from HMM state occupancy statistics \cite{Duration}. During the speech synthesis process, the most likely state sequence is first determined according to the duration model. The most likely feature vector sequence associated to that state sequence is then generated, as described in \cite{Generation}. Finally, these feature vectors are fed into a vocoder to produce the speech signal.

\begin{figure}[!ht]
  \centering
  \includegraphics[width=0.5\textwidth]{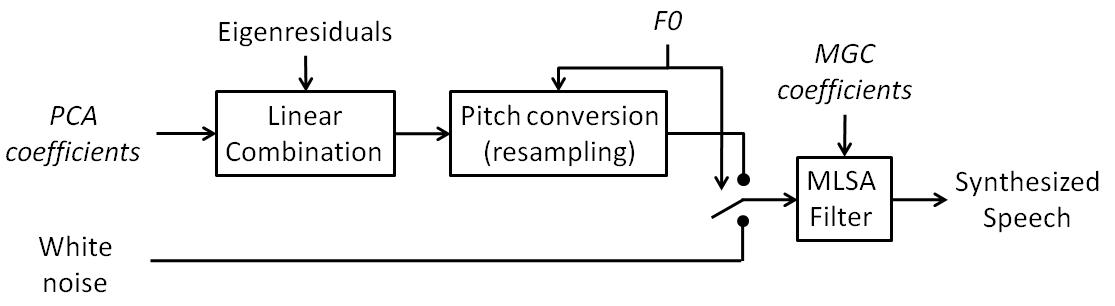}
  \caption{Vocoder framework used during the synthesis stage. Inputs are the PCA, F0 and MGC coefficients generated by the HMMs.}
  \label{fig:vocoder}
\end{figure}

The vocoder workflow is depicted in Figure \ref{fig:vocoder}. The generated $F_0$ value commands the voiced/unvoiced decision. During unvoiced frames, white noise is used. On the opposite, the voiced frames are constructed according to the synthesized PCA coefficients. A first version is obtained by linear combination with the eigenresiduals extracted as detailed in Section \ref{sec:Eigenres}. Since this version is size-normalized, a conversion towards the target pitch is required. As justified in Section \ref{ssec:dataset}, this can be achieved by resampling. The choice we made during the normalization of a sufficiently low pitch (cf. Equation \ref{eq:pitch}) is now clearly understood as a constraint for avoiding the emergence of energy holes at high frequencies. Frames are then overlap-added so as to obtain the excitation signal. The so-called Mel Log Spectrum Approximation (MLSA) filter, based on the generated MGC coefficients, is finally used to get the synthesized speech signal.

\subsection{Subjective test results}

\begin{figure}[!ht]
  \centering
  \includegraphics[width=0.45\textwidth]{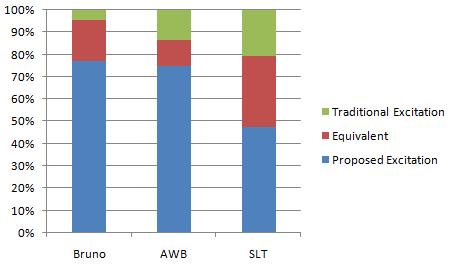}
  \caption{Results of the preference test.}
  \label{fig:MOS}
\end{figure}

Three voices were evaluated: Bruno (French male), kindly provided by Acapela Group, AWB (Scottish male) and SLT (US female) from the CMU ARCTIC database available in \cite{CMU}. The training set had a duration of about 50 min for AWB and SLT, and 2 h for Bruno and was composed of phonetically balanced utterances sampled at 16 kHz. The subjective test was submitted to 20 non-professional listeners. It consisted of 4 synthesized sentences of about 7 seconds per speaker. For each sentence, two versions were presented (using either the traditional or proposed excitation) and the subjects were asked to vote for the one they preferred (if any). Averaged results are shown in Figure \ref{fig:MOS}. A considerable gain through the use of our excitation is reached on the male voices. Although still present, this advantage turns out to be less dominant for the female speaker.

\vspace{-9pt}
\section{Conclusions and future works}\label{sec:conclu}

This paper proposed a new model of excitation for parametric speech synthesis. This model relies on the decomposition of pitch-synchronous excitation frames on a limited number (typically around 20-30) of eigenresiduals, obtained on a relatively small dataset (20 min are generally sufficient). This approach was tested on HMM-based synthesis by enhancing the source modeling with PCA-based coefficients. A substantial improvement was reported on male as well as female (even though less pronounced) speakers, while the system footprint still remains around 1Mb.

As future works, we plan to investigate the following directions: 

\begin{itemize}

\item
One of the main problems encountered in voice conversion could be alleviated by the proposed approach. Indeed, while such systems achieve their main goal, i.e the converted voice is recognized as being uttered by the target speaker, they suffer from a poor quality in the delivered speech \cite{Yannis}. Replacing the traditional pulse excitation by the use of eigenresiduals (beforehand computed for the target speaker) should also undoubtedly lead to significant improvements.

\item
Since the glottal source could differ with the phonetic context, analyzing the effect on a preliminary stage of clustering on the eigenresiduals would be worthwhile. More precisely, the HMM-based speech synthesizer makes use of binary decision trees achieving a Context-Oriented Clustering. We therefore plan to apply such a Principal Component Analysis for different nodes of these trees, and observe whether a significant difference in the eigenresidual waveform is noticeable, and if this way of proceeding leads to an improvement in the quality of the synthesized speech (sufficiently important for the increase of complexity).

\item
Although we proposed the use of a Principal Component Analysis, other data mining methods (possibly derived from the functional PCA literature, \cite{Silverman}) could be efficiently employed to extract a suitable representation from the large dataset of normalized GCI-centered residual frames (obtained as described in Section \ref{ssec:dataset}).

\item
Finally, it would certainly be very interesting to compare the proposed approach with other techniques of excitation modeling, such as STRAIGHT \cite{STRAIGHT}, the mixed excitation \cite{Yoshimura},\cite{Maia}, or based on the Liljencrant-Fant model \cite{Cabral2}. Although all these approaches reported a relative improvement with regard to the traditional pulse excitation, no comparison is available yet, since authors worked with different synthesis frameworks and with different databases. 

\end{itemize}

\section{Acknowledgments}\label{sec:Acknowledgments}

Thomas Drugman is supported by the ``Fonds National de la Recherche
Scientifique'' (FNRS). Authors also would like to thank the reviewers for their helpful feedback.



\end{document}